\documentclass{du-journals}
\journal{Iranian Journal of Astronomy and Astrophysics}
\title{Review of Image Processing Methods in Solar Photospheric Data Analyzes}
\subtitle{Photospheric Image Processing Methods}
\author[1]{Mohsen Javaherian}
\address[1]{Research Institute for Astronomy and Astrophysics of Maragha (RIAAM), University of Maragheh, 55136-553, Maragheh, Iran; email: javaherian@maragheh.ac.ir}
\author[2]{Zahra Eskandari}
\address[2]{Department of Geography and Rural Planning, University of Zanjan, University Blvd., 45371-38791, Zanjan, Islamic Republic of Iran; email: z.eskandari@znu.ac.ir}
\begin{document}
\begin{abstract}
With the exponential growth in data volume, especially in recent decades, the demand for data processing has surged across all scientific fields. Within astronomical datasets, the combination of solar space missions and ground-based telescopes has yielded high spatial and temporal resolutions for observing the Sun, thus fueling an increase in the utilization of automatic image processing approaches. Image processing methodologies play a pivotal role in analyzing solar data, a critical component in comprehending the Sun's behavior and its influence on Earth. This paper provides an overview of the utilization of diverse processing techniques applied to images captured from the solar photosphere. The introduction of our manuscript furnishes a description of the solar photosphere along with its primary characteristics. Subsequently, we endeavor to outline the significance of preprocessing photospheric images, a crucial prerequisite before engaging in any form of analysis. The subsequent section delves into an examination of numerous reputable sources that have employed image processing methodologies in their research pertaining to the Sun's surface. This section also encompasses discussions concerning recent advancements in image processing techniques for solar data analysis and their potential implications for future solar research. The final section deliberates on post-processing procedures as supplementary steps that are essential for deriving meaningful results from raw data. Effectively, this paper imparts vital information, offering concise explanations regarding the Sun's surface, the application of image processing techniques to various types of photospheric images, indispensable image preprocessing stages, and post-processing procedures aimed at transforming raw data into coherent and comprehensive insights.
\end{abstract}


\section{Introduction}
The solar photosphere is the visible surface of the Sun and is the layer from which most of the Sun's radiation is emitted. It is the layer that we observe when we look at the Sun using a telescope or with the naked eye using a solar filter. The photosphere is a thin layer of gas that has a thickness of only a few hundred kilometers \cite{Foukal2004,Kuhn2016solar}. It is a highly dynamic and complex layer, with constant convective motions that give rise to granulation, which are small cellular structures \cite{Nordlund2009}. The photosphere of the Sun is covered by granulation, which consists of irregularly shaped convective cells. These cells are constantly forming and disappearing in a turbulent manner. The bright center of each cell contains hot, rising plasma that flows horizontally at a speed of 0.5 to 1.5 km s$^{-1}$. The boundaries, which are about 0.3 Mm wide and are referred to as intergranular lanes, are dark and represent cooler, falling material. Granules typically have a diameter of 1 Mm, but their size can range from 0.3 to 2 Mm. The turnover time for a granule with a diameter of 1 Mm and a speed of 1 km s$^{-1}$ is about 1000 sec or approximately 20 min. Granules have intensity contrasts of 5 to 15 per cent in white light, depending on the resolution, or up to 32 per cent in the near UV \cite{Hirzberger2010}. The lifetime of a granule is typically about 5 to 10 min, but it can range from 1 to 20 min. Larger granules with smaller random horizontal velocities tend to have longer lifetimes. The evolution of a granule is strongly influenced by its environment, including the nearby magnetic field and its location within meso- and super-granules \cite{Title1989}. Granules move with both meso- and super-granules. Granules are born from the merging of two smaller granules or the splitting of a larger one, such as an exploding granule \cite{Priest2014}. The characteristic feature of this layer is sunspots, which have been concerned by many ancients for a long time. These large-scale magnetic structures are dark and cool regions on the Sun's surface that are associated with intense magnetic fields \cite{Berdyugina2005,Livingston2019sunspots} and evolution in active regions (ARs) \cite{Gesztelyi2015}. This layer is also origin of lots of phenomena related to the small-scale magnetic fields \cite{deWijn2009} that can be observed on the photosphere's surface. The concentration of magnetic fields as visible-light signatures can be studied in one of the following classes: faculae, filigree, facular bright points (BPs), and network BPs \cite{Berger1995}. The other fundamental magnetohydrodynamic event is solar granulation which is explained as a process of advection-fragmentation that occurs in the upper layers of the convection zone. The reader can refer to \cite{Leighton1963}, \cite{Bray1997solar}, \cite{Title1998granul}, \cite{Muller1999}, \cite{Shine2000superg}, \cite{Nesis2006}, and \cite{Rincon2018sun} for more details and review about granulation, mesogranulation, and supergranulation dynamical processes in the Sun.

Temperature is a measure of the average kinetic energy of the gas particles in the photosphere. Different theories explain that the temperature of the photosphere typically decreases with distance from the sun's core, with a range of around 4500 to 6000 Kelvin. Pressure is a measure of the force exerted by the gas particles on their surroundings and is essential for understanding the dynamics of the photosphere. The pressure in the photosphere varies with height and temperature, with a range of around 0.1 to 100 Pa. The density is relatively low compared to the layers above it, with a value of around 10$^{-7}$ kg.m$^{-3}$. Magnetic fields in the photosphere is typically weaker than in the layers above it and are also critical since they can influence the behavior of the plasma \cite{Solanki2013sun}, giving rise to phenomena such as sunspots \cite{Borrero2011}, CMEs \cite{Chen2011,Webb2012}, and flares \cite{Shibata2011}. The strong magnetic fields in the solar photosphere are concentrated into small magnetic elements or intense flux tubes. These structures typically have field strengths of 1 kG, fluxes of $3 \times 10^9$ Wb ($3 \times 10^{17}$ Mx), and diameters of 100 km \cite{Stenflo1973}. As the flux tubes rise through the photosphere, their magnetic field strength decreases, from 1500 to 1700 G in the deep photosphere to 1000 to 1200 G in the middle photosphere and 200 to 500 G at the temperature minimum. The ratio of plasma pressure to magnetic pressure ($\beta$) is small in the tubes, ranging from 0.2 to 0.4, indicating that the magnetic field dominates the plasma. In the neighboring photosphere, $\beta$ is greater than 1, whereas in the chromosphere and corona, where the magnetic flux spreads, $\beta$ is much less than 1 in ARs and is typically less than or equal to 1 elsewhere \cite{Priest2014}.

Indeed, magnetohydrodynamics (MHD) theory can help to determine the range of parameters of the Sun \cite{Priest2007magnetic,Priest2014}, however, the exact measurement of the physical parameters of the photosphere is a challenging task since they cannot be measured directly. Instead, scientists use various indirect methods to infer these parameters. For example, the temperature of the photosphere can be estimated by analyzing the spectrum of the light emitted by the Sun. Pressure can be inferred from the observed granulation patterns, and magnetic fields can be measured using a technique called Zeeman splitting, which splits spectral lines in the presence of a magnetic field. There are also some challenges associated with measuring photospheric parameters, including the effects of atmospheric turbulence and instrument calibration. Atmospheric turbulence can cause distortions in the images, making it difficult to obtain accurate measurements, and instrument calibration is essential to ensure that the measurements are accurate and reliable. In recent years, there have been significant advances in photospheric observations, including the development of new instruments and techniques that allow for more precise measurements of the physical parameters of the photosphere. These advances have led to a better understanding of the photosphere's behavior and its role in the Sun's activity cycle. Further research in this area will continue to improve our understanding of the Sun's behavior and its impact on our planet \cite{Priest2007magnetic,Eddy2009,Zbiciak2023}.

\section{Significance of Pre-processing Procedures in Solar Images}

In an excellent article presented by \cite{Aschwanden2010a} explained that the image pre-processing procedures in solar image analysis are essential for successful automated feature detection algorithms. These procedures aim to remove unwanted features, correct artifacts, enhance relevant structures, and standardize images for further analysis. Here, we outline key aspects of these procedures: $\bullet$ Instrumental Effects: Solar images acquired by space-borne telescopes often contain artifacts such as "dark pixels" and "hot pixels" due to the imperfections of charge-coupled devices (CCDs). These artifacts need to be corrected through methods like dark current subtraction or pedestal subtraction to prevent false detections. Also, issues like vignetting, silhouetting, and instrument-induced background must be accounted for to avoid incorrect measurements. $\bullet$ Non-Solar Features: To isolate solar phenomena, non-solar features like cosmic rays or high-energy particle hits must be suppressed. Despiking and destreaking algorithms, along with mathematical morphology filters, help remove such artifacts. These methods help eliminate unwanted signals and enhance the accuracy of feature detection. $\bullet$ Image Resolution: Insufficient resolution can lead to inaccurate measurements. The spatial resolution of different instruments must be characterized, and images may need to be deconvolved or resampled to a consistent resolution for reliable automated feature detection. $\bullet$ Image Background: Background subtraction is crucial for feature detection. The challenge lies in defining suitable background models, considering both feature-unrelated structures and instrument-related background flux. Neglecting this step can lead to inaccurate feature measurements and analyses. $\bullet$ Coordinate System: Images from different sources need to be coaligned in identical coordinate systems. This ensures accurate feature tracking and quantitative analysis across various datasets. Various coordinate transformations and coalignment methods are used to align images effectively. $\bullet$ Image Filtering: Image filters are employed to enhance specific features while suppressing noise and unrelated structures. Fourier and wavelet filtering techniques are common for solar image enhancement. Non-linear filters like unsharp masking can significantly enhance morphology and geometry, aiding feature detection. $\bullet$ Image Restoration and Reconstruction: Seeing effects in ground-based solar images can distort features. Techniques like speckle imaging and phase diversity are used for restoration. In radio and hard X-ray imaging, Fourier-type imaging is employed to reconstruct images, accounting for wavelength-dependent effects.

For preprocessing photospheric images, some points must be carefully fulfilled. First of all, the noise reduction procedure is applied to the data, if it needs. In some data sets taken by different telescopes, there are routines to increase the signal-to-noise ratio (SNR) and denoise the recorded images. Setting a particular value for SNR helps to achieve a satisfactory level of restoration and avoiding the amplification of unwanted noise. We can point to the flat field correction and dark current subtraction as steps of preprocessing procedures. If the purpose of a study is tracking features in sequential frames, two main tasks must be fulfilled before any type of processing:

\textbf{Derotation} A crucial data processing technique used in solar astronomy is derotation of solar images to compensate for the Sun's rotation when observing solar features over an extended period. As the Sun rotates, different parts of its surface come into view at different times, leading to apparent motion of solar features in observational data. Derotation helps align solar images taken at different times to a common reference frame, allowing researchers to study and track solar phenomena with improved temporal coherence and accuracy. For more specialized explanations of solar differential rotation and technique of derotation, see \cite{Carrington1859,Howard1976,Hathaway2022} and \cite{Starck2006,Viladrich2020}, respectively.

Here's some points about the derotation process and its significance in solar image analysis: $\bullet$ Sun's Rotation: The Sun is a gaseous, rotating celestial body. Its equator rotates at a faster rate than its poles. As a result, the solar surface completes one full rotation approximately every 25 days at the equator and around 35 days at the poles. This rotation causes solar features, such as sunspots, prominences, and filaments, to appear to move across the solar disk when observed over several hours or days. $\bullet$ Need for Derotation: When studying solar phenomena that evolve over longer periods, it is essential to compensate for the Sun's rotation. Without derotation, the apparent motion of features would introduce significant errors in measuring their properties, trajectories, and changes over time. Derotation is particularly important for long-term observations and time-series analysis of solar activity. $\bullet$ Derotation Techniques: There are various methods to derotate solar images, depending on the data and the specific study objectives. One common approach is to use solar tracking software, which precisely measures the Sun's position and orientation at different times and then applies an inverse rotation to align the images to a common reference frame. This method requires accurate tracking of the solar limb or other features on the solar disk. $\bullet$ Differential Rotation: The Sun exhibits differential rotation, meaning that different latitudinal bands rotate at slightly different rates. This effect further complicates the derotation process. More sophisticated derotation techniques take into account the differential rotation to ensure accurate alignment of features over time. $\bullet$ Advantages of Derotation: Derotation allows solar researchers to create time-lapse movies or image sequences that show the evolution of solar phenomena with a stable reference frame. This stability enhances the ability to analyze solar activities, track solar features, and study long-term trends in solar behavior. $\bullet$ Applications: Derotation is used in various solar studies, such as tracking sunspot groups and magnetic fields, studying the evolution of solar flares and prominences, and investigating solar surface dynamics and large-scale motions. It is also instrumental in studying the solar cycle and long-term variations in solar activity. $\bullet$ Limitations: While derotation greatly improves the coherence of solar image sequences, it may introduce some artifacts or errors, especially in the presence of atmospheric turbulence or instrumental imperfections. Advanced image processing techniques and careful data calibration are often used to mitigate these issues. For example, see \cite{Priest2007magnetic},  for further details.

\textbf{Subsoinic filter} The term "subsonic $k - \omega$ filter" in the context of solar physics refers to a data processing technique used to modify or analyze 5-minute photospheric oscillations. These oscillations are waves that propagate through the Sun's photosphere with a characteristic period of around 5 minutes correspond to a cutoff velocity of 7 km s$^{-1}$. These oscillations are commonly referred to as solar p-modes (pressure modes) and are associated with the Sun's acoustic and gravity waves \cite{Sobotka1999,Sobotka2001}.

The subsonic filter is applied to time-series data (cube), such as Doppler velocity or intensity observations of the solar photosphere, obtained from ground-based or space-based instruments. The main goal of this filtering technique is to remove or suppress high-frequency noise and unwanted signals, revealing the underlying 5-minute oscillation signal more clearly. This allows researchers to study the properties and behavior of these oscillations in greater detail. The subsonic filter typically involves the following steps: $\bullet$ Fourier Transform: The time-series data is transformed from the time domain to the frequency domain using a Fourier transform. This transformation allows the data to be analyzed in terms of its constituent frequency components. $\bullet$ Frequency Filtering: The subsonic filter focuses on low-frequency components corresponding to the 5-minute p-mode oscillations. It may involve the application of various mathematical techniques, such as bandpass filtering or wavelet analysis, to isolate the desired frequency range. $\bullet$ Removal of High-frequency Noise: High-frequency noise and unwanted signals are removed or significantly attenuated from the data using the filter. This process helps to enhance the visibility of the 5-minute oscillation signal. $\bullet$ Inverse Fourier Transform: After the filtering process, an inverse Fourier transform is applied to convert the data back to the time domain. The filtered time-series data now contains the modified 5-minute photospheric oscillation signal with reduced noise and interference.

\section{Image Processing Methods}

As known, "image processing methods" refer to a variety of techniques used to manipulate and analyze digital images. These methods can be broadly categorized into four groups: image enhancement, image segmentation, object recognition, and image compression. Image enhancement approaches aim to improve the quality of an image by adjusting its brightness, contrast, color balance, and sharpness. Common techniques used for image enhancement include histogram equalization, contrast stretching, and spatial filtering. Image segmentation methods involve dividing an image into multiple regions or segments based on their characteristics such as color, texture, or intensity. Commonly, segmentation process is fulfilled by region-based and/or edge-based procedures. The technique of segmentation is useful for object detection and tracking, image compression, and data analysis. Some known image segmentation methods include region growing, edge detection, thresholding, and clustering. Object recognition is used to identify and classify objects within an image. It involves a series of steps, such as feature extraction, feature matching, and classification. Object recognition techniques are often used in computer vision applications, including robotics, autonomous vehicles, and facial recognition. Some popular object recognition methods include template matching, neural networks, and support vector machines. Image compression involves reducing the size of an image by removing redundant or irrelevant information without significant loss of image quality. Image compression is essential for storing and transmitting large volumes of image data efficiently. Popular image compression methods include JPEG, PNG, and GIF. The interested reader can refer to \cite{Szeliski2010}, \cite{Sayood2017}, \cite{Pratt2018}, and \cite{Gonzalez2018}.

\begin{figure}
\centerline{\includegraphics[width=15cm]{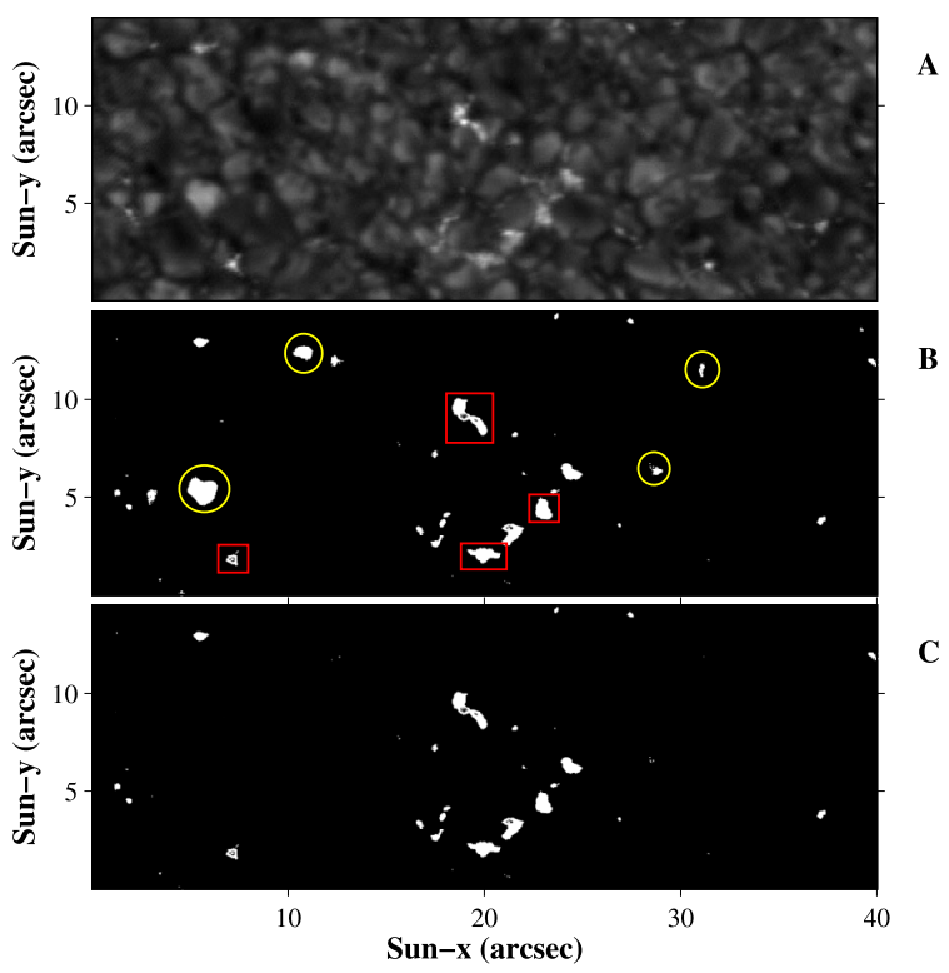}}
\caption[]{On 9 June 2009 at 14:44:03 UT, a 214 nm image was acquired using Sunrise/SuFI (A). The outcome of the region-growing function is displayed in (B), with highlighted red rectangular boxes marking samples of network BPs and yellow circles denoting non-BPs. For final recognition of BPs, it needs a supervised classification technique such as support vector machine (SVM). The results of the SVM classifier are shown as BPs in (C). For further information, see \cite{Javaherian2014}. Image reproduced with permission from \cite{Javaherian2014}, copyright by Springer.}
\label{BPs}
\end{figure}

First of all, we must able to segregate a solar feature from background using digital image processing methods. So, in order to define features in photospheric images, a threshold is typically required to separate the feature-related signal from the background. However, since features can display a wide range of morphological and geometric structures, the background structure can be equally diverse and not easily separated from feature of interest. A simple flux threshold may capture bright structures such as the BPs, but may miss fainter network BPs. On the other hand, setting the threshold very low to include faint features may result less brighter features than individual BPs (see the caption of Figure \ref{BPs}). Thus, a simple threshold-based background subtraction method is generally not sufficient for BP detection, and more sophisticated image enhancement and filtering approaches are required to extract the feature of interest from background \cite{Aschwanden2010a}. We should mention that the image processing methods that we explain here, it would lead to automatic recognition of photospheric features or not. Roughly speaking, in so many times image processing methods are primary steps of automatic recognition procedure. The reader can refer to \cite{Aschwanden2010a}, \cite{Aschwanden2010b}, \cite{Arish2016}, \cite{Tajik2023}, and \cite{AsensioRamos2023} for review of automatic recognition and machine learning methods in solar physics.

In 1986, Roudier and Muller conducted an analysis of the structure of solar granulation using computer-processed broad-band images obtained at the Pic-du-Midi Observatory. To enhance the granulation spatial scale and filter out large-scale variations, the authors applied a Fourier-based procedure on the images. They then extracted the granular structures by applying a single threshold \cite{Roudier1986}. Hirzberger and colleagues (1997, 1999a,b) employed a threshold on band-pass filtered images to identify granules in their series of papers \cite{Hirzberger1997}, \cite{Hirzberger1999a}, and \cite{Hirzberger1999b}. They defined the granular cells as areas that included the related granule and half of the surrounding intergranular lane. The chosen algorithm for detecting granules greatly impacts their automatic definition. One approach involved using a Laplacian operator to identify the inflection points in the intensity distribution of the images, which were then used to define the borders of the granules. To enhance the contrast of the granular substructures, they smoothed images before applying the operator. In paper \cite{Title1989}, the authors tested using single thresholding to identify granules, but found that this method was too sensitive to low spatial frequency intensity variations and the specific spatial filter used. Instead, they developed two procedures to detect granules: a center-finding approach that explored the local neighborhood to locate local maxima associated with granules, and a lane-finding method that surrounded the granules with a boundary lane. These techniques were used to identify, measure, and analyze the properties of solar granulation observed in movies captured by the Solar Optical Universal Polarimeter during the Spacelab 2 mission \cite{Title1989}.

The paper \cite{Berrilli2005} describes two main methods for detecting granules. The first is based on a multi-scale Laplacian-of-Gaussian (LoG) operator. The LoG operator is a filter that enhances structures of a specific size in an image. By applying the LoG operator at multiple scales, the filter can detect structures of different sizes, including granules. The authors apply the LoG operator to intensity and Doppler images of the photosphere to extract compact structures. The multi-scale LoG operator has several advantages over other approaches for granule detection. It is a non-subjective approach, meaning that it does not require manual input or tuning of parameters. It also does not require assumptions about the size or shape of granules, which can vary depending on the conditions of the solar atmosphere. The multi-scale LoG operator can detect granules of different sizes and shapes, as well as other compact structures in the photosphere. The paper \cite{Berrilli2005} also provides a detailed description of how the multi-scale LoG operator is applied to solar images. The authors first apply a Gaussian filter to the image to suppress noise. They then compute the Laplacian of the Gaussian at multiple scales, using a scale-space representation of the image. The scale-space representation is a pyramid of images, where each level corresponds to a different scale. The Laplacian of the Gaussian is computed at each level of the pyramid, and the resulting images are combined to obtain a final image that highlights structures of different sizes.

One of the applicable methods in segmentation and identification methods has been firstly introduced by \cite{Bovelet2001} and then employed for extraction of granules and BPs from images \cite{Bovelet2007}. They examined small-scale magnetic flux concentrations on the Sun using high-resolution G-band images and compared them with Ca ii H enhancements. Their identification algorithm of G-band structures operates in four steps: Firstly, a set of equidistant detection levels is used to create a pattern of primary "cells". Secondly, the intrinsic intensity profile of each cell is normalized to its brightest pixel. Thirdly, the cell sizes are shrunk by a unitary single-intensity clip. Finally, features in contact at an appropriate reference level are merged by removing the respective common dividing lines. Optionally, adjoining structures may be excluded from this merging process, referring to the parameterized number and intensity of those pixels where enveloping feature contours overlap. All these steps were included in the multiple level tracking
(MLT) algorithm.  Magnetic intergranular structures (MIgS) are then selected from the intergranular structures (IgS) pattern based on their local Ca ii H contrast and their mean G-band-to-continuum brightness ratio. Out of 970 small-sized G-band IgS, 45\% were found to be MIgS and co-spatial with isolated locations of Ca ii H excess. These MIgS may be twice as frequent as the known G-band BPs.

\begin{figure}
\centerline{\includegraphics[width=9cm]{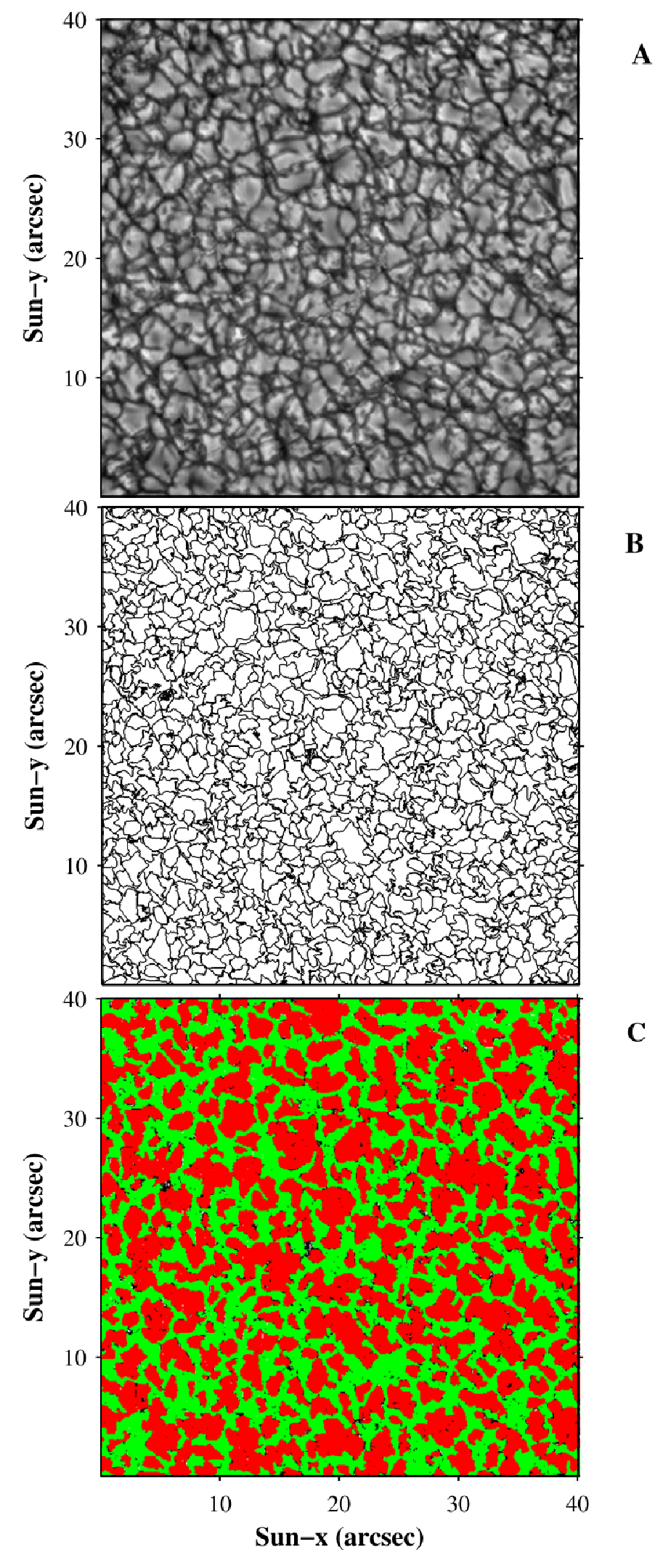}}
\caption[]{A Sunrise/IMaX image taken at 5250.02 nm on 9 June 2009 at 14:27:14 UT is depicted in (A). Utilizing the mean shift procedure, the image has been segmented, as demonstrated in (B). The results of the SVM classifier are displayed in (C). The color-coded regions indicate granules in red and non-granular regions in green. For details, refer to \cite{Javaherian2014}. Image reproduced with permission from \cite{Javaherian2014}, copyright by Springer.}
\label{Gs}
\end{figure}

A new promising automatic code is improved by \cite{Javaherian2014} for photospheric image segmentation and feature extraction. This approach is performed based on automatically detecting photospheric features (BPs and granules) from ultraviolet (UV) radiation using a feature-based classifier. This method employs images of the quiet Sun (QS) at 214 nm and 525 nm captured by Sunrise balloon-borne solar observatory \cite{Barthol2011,Solanki2010} using Sunrise Filter Imager (SuFI)\cite{Gandorfer2011} and Imaging Magnetograph eXperiment (IMaX)\cite{MartinezPillet2011} telescope, respectively, on 9 June 2009 (see also \cite{Berkefeld2011} for more information). The region growing and mean shift procedures are utilized to segment BPs and granules, respectively, and calculate Zernike moments of each region. The function Zernike polynomials is applied to the segmented images to extract scale-, translation-, and rotation-invariant moments. The region growing segmentation has applications in various fields, including computer vision, object tracking, and image analysis. Also, the Mean Shift procedure is a powerful technique for image segmentation, especially when dealing with images that contain a variety of textures and colors without clear boundaries. So, we decided to add the explanations to discuss about about these two segmentations in details:

Region growing is a common technique used in image processing and computer vision for image segmentation. The main idea behind region growing is to group pixels or image elements that share certain common characteristics, such as intensity values or color, into meaningful regions. This process helps to segment an image into distinct regions that correspond to objects or structures of interest.

The region growing algorithm starts with one or more seed points that are chosen from the image. These seed points are typically selected based on some criteria, such as user input or initial processing steps. Once the seed points are chosen, the algorithm iteratively grows the regions by comparing the properties of neighboring pixels or elements to the properties of the current region. Here's a step-by-step explanation of the region growing process: $\bullet$ Seed Point Selection: One or more seed points are chosen as starting points for the segmentation process. These seed points are often manually selected based on prior knowledge or user input. $\bullet$ Initialization: The properties of the seed points, such as intensity values, color values, or texture characteristics, are used to define the initial region. $\bullet$ Neighbor Checking: For each pixel or element adjacent to the current region, its properties are compared to the properties of the current region. If the properties of the neighboring pixel are similar enough to those of the region, it is added to the region. $\bullet$ Growing Criteria: The similarity or dissimilarity criteria that determine whether a neighboring pixel is added to the region can vary depending on the application. Common criteria include comparing intensity values, color differences, gradient values, or texture features. $\bullet$ Connectivity Consideration: Depending on the connectivity criteria chosen, neighboring pixels or elements can be 4-connected (horizontally and vertically adjacent) or 8-connected (including diagonally adjacent). $\bullet$ Iteration: The process of comparing neighboring pixels to the region and adding them if they meet the criteria continues iteratively until no more pixels can be added. $\bullet$ Termination: The region growing process terminates when no more pixels meet the criteria for inclusion in the region, or when a predefined stopping condition is met (e.g., a maximum region size). $\bullet$ Post-processing: Depending on the application and the quality of the segmentation, post-processing steps may be applied to refine the segmented regions. This could include removing small isolated regions, merging regions that are too small, or splitting regions that are too large.

The Mean Shift procedure is a popular technique used in image segmentation, a process of partitioning an image into meaningful regions based on certain characteristics or criteria. Mean Shift is particularly effective in situations where there are no well-defined edges or color boundaries in the image, making it suitable for segmenting regions with varying textures, colors, and gradients. Here's how the Mean Shift procedure for image segmentation works: $\bullet$ Kernel Density Estimation: The Mean Shift algorithm starts with a set of data points in the image, which could be pixel values in the feature space (e.g., color, intensity, texture). Each data point is associated with a kernel, which is a mathematical function that assigns weights to neighboring points based on their distance. $\bullet$ Centroid Initialization: A window (also called a kernel or search window) is placed around each data point. This window defines the region from which neighboring points will influence the current data point. Initially, each data point's window is centered at itself. $\bullet$ Shift to Local Maximum: For each data point, the mean shift procedure is applied. The idea is to compute the mean shift vector, which indicates the direction in which the density of neighboring points is highest. In other words, it points towards the direction of the local maximum of the kernel density estimation. $\bullet$ Updating the Data Point's Position: The data point is then shifted in the direction of the mean shift vector. This essentially moves the data point towards the region of higher density, which often corresponds to a cluster of similar points in the image. $\bullet$ Convergence and Clustering: Steps 3 and 4 are repeated iteratively until the data point converges to a local maximum. As the data point moves, it gathers other nearby data points that are part of the same cluster. When convergence is reached, the final position of the data point becomes the representative or centroid of the cluster. $\bullet$ Labeling and Segmentation: After convergence, all data points that have the same centroid are considered part of the same cluster. This process effectively segments the image into distinct regions based on the clusters formed during the mean shift procedure.

Region growing is a versatile technique that can be adapted to various image segmentation tasks, including both grayscale and color images. However, it also has some limitations, such as sensitivity to seed point selection and the potential to get stuck in local minima if the similarity criteria are not well-defined. Nonetheless, when used appropriately and with careful parameter tuning, region growing can be an effective method for segmenting images into coherent and meaningful regions (e.g., see \cite{FU1981,GAMBOTTO1993,Hojjatoleslami1998,Gonzalez2018}). On the other hand, the mean Shift offers several advantages, including its ability to handle non-linear boundaries and adapt to different shapes and sizes of regions. However, it might require careful tuning of parameters like the bandwidth of the kernel, which determines the size of the search window and influences the scale of the segmentation (e.g., see \cite{Comaniciu2002,Hong2007,Li2010Mean}).

\begin{figure}
\centerline{\includegraphics[width=15.5cm]{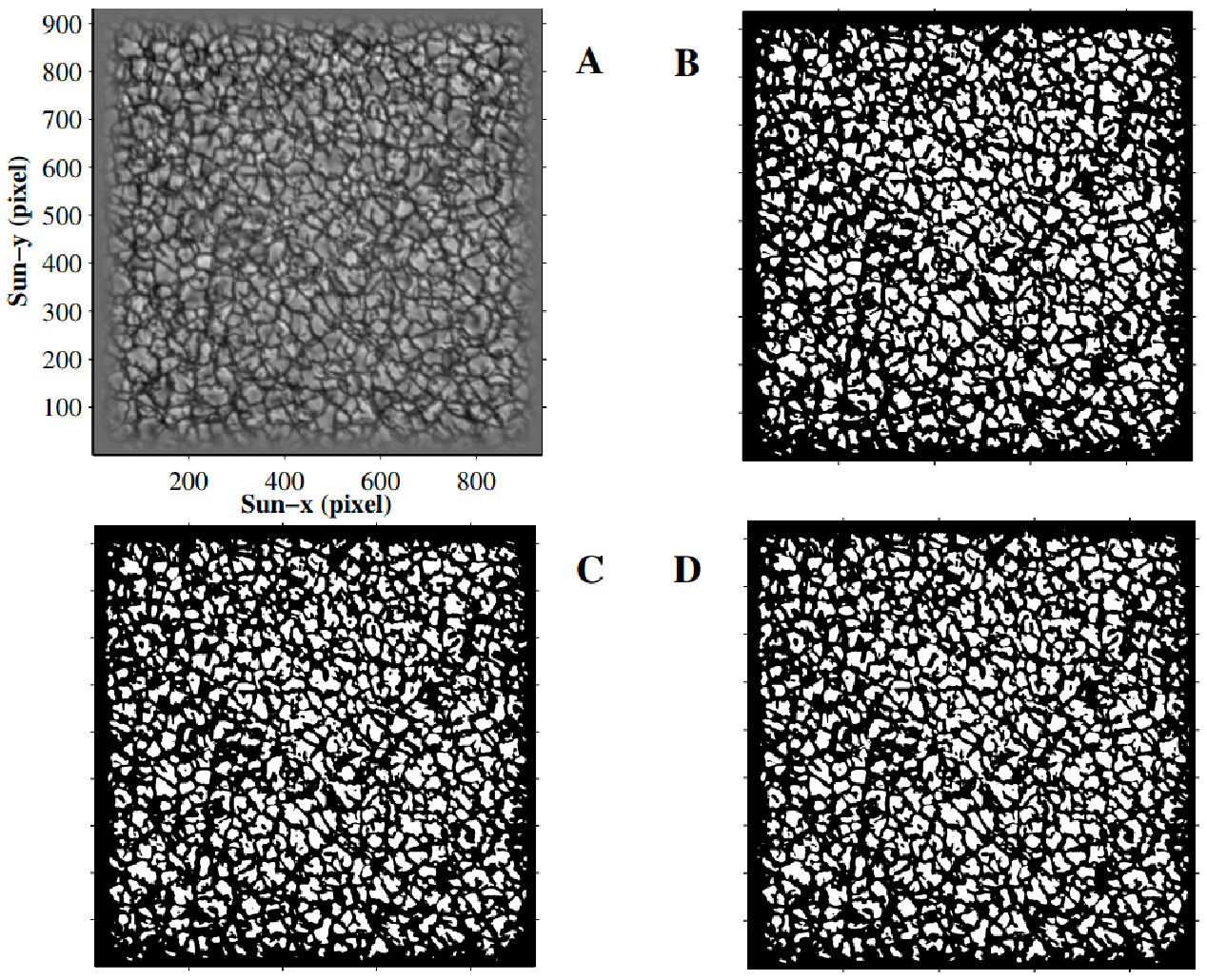}}
\caption[]{The Sunrise/IMaX captured an image at 14:16:00 UT on 9 June 2009 (A). Subsequently, the segmentation process was applied using the c-means technique (B), the k-means approach (C), and the FCM procedure (D). The reader can refer to \cite{Yousefzadeh2015}. Image reproduced with permission from \cite{Yousefzadeh2015}, copyright by IJAA.}
\label{FCM}
\end{figure}

\cite{Yousefzadeh2015} employed three different clustering methods, namely c-means, k-means, and fuzzy c-means (FCM) algorithms, to segment solar ultra-violet (UV) images. The methods are applied to a sequence of photospheric observations of QS at 525 nm taken by the Sunrise $/$ IMaX on June 9, 2009 UT. The original image and segmentation results of exerted algorithms are shown in Figure \ref{FCM}. While these algorithms produce slightly different results in terms of extracting physical parameters such as filling factors, brightness fluctuations, and size distributions from the images, the FCM algorithm yields a mean granule size of approximately 1.8 arcsec$^2$ (0.85 Mm$^2$). They found that the smaller granules with sizes less than 2.8 arcsec$^2$ exhibit a wide range of brightness, whereas larger granules approach a more uniform value.

Among data related to the photosphere, lots of image processing methods especially in segmentation supergranular cells and magnetic patches were developed for magnetogram, dopplergram, and continuum image data sets. There are several techniques used to calculate the flow velocities to extract supergranular cell regions.

One of the methods for estimating the proper motions of tracers observed in successive images of photosphere is called local correlation tracking (LCT), which was introduced by \cite{November1988} and later refined by \cite{Simon2001}. The central idea behind the various LCT schemas in use today is to calculate proper motions of intensity features in successive images, such as G-band filtergrams, H$\alpha$ images, or photospheric magnetograms, separated in time by a given cadence. This is typically accomplished by either minimizing an error function between subregions of the sequential images or maximizing a cross-correlation function. In this method, a mask of $3 \times 3$ is employed to shift images at nine integer-pixel spatial lags, including one null lag and eight toward nearest-neighbor pixels. At each shift, a cross-correlation function between pixels in consecutive frames is computed. The cross-correlation is obtained by integrating the product of a Gaussian windowing function with the smoothed before and after images at the given shift. To achieve subpixel resolution, they use biquadratic interpolation on the nine-point cross-correlation distribution to determine the shift that corresponds to the maximum correlation. The velocity is then calculated by multiplying the shift by the pixel size and dividing by the time interval (cadence) between the two images. There is another approach for minimizing error introduced by \cite{Berger1998} which involves minimizing the root-mean-square (rms) error between corresponding subimages, or tiles, as they are commonly known. In this method, each subimage from one image is shifted until it is most closely aligned with the corresponding subimage from the other image. Once the optimal shifts have been determined, the pattern of subimage shifts reveals the motion that occurred over the time interval between the two images (cadence). The overall velocity pattern can then be determined by interpolating between the shifts of the tiles. Nowadays, LCT is widely used for various types of applications \cite{Krijger2003,Vargas2010,Verma2013,Bai2022,Li2023}.

Fast Local Correlation Tracking (FLCT) is a method for estimating the correlation tracking error in solar differential rotation measurements. The method was developed by \cite{Fisher1999} and employs Fourier correlation to obtain the velocity field of solar features on the solar surface. To use FLCT, the first step is to apply a Gaussian filter to the images to be correlated. The width of the Gaussian filter is determined by the user and corresponds to a spatial scale over which the flow is to be tracked. For each pixel in the image array, two images are multiplied by Gaussian filters centered at that pixel. The resulting images are then cropped to remove the insignificant parts of the images, thereby reducing the image size and increasing computational speed. Next, the cross-correlation function between the two cropped images is computed using standard fast Fourier transform (FFT) techniques. The shifts in x and y that maximize the cross-correlation function are then found using cubic-convolution interpolation. These shifts are used along with the time difference between the images to determine the velocities of the intensity features along the solar surface \cite{Fisher2008}. The FLCT method is similar to other LCT methods, but it uses a Gaussian filter instead of a rectangular filter to weight the data. This has the advantage of providing smoother and more continuous velocity fields. Additionally, FLCT is computationally efficient, making it well-suited for analyzing large datasets. Although FLCT was originally developed for use in solar differential rotation measurements, it has also been used in other applications such as tracking the motion of granules in the solar photosphere and tracking the motion of clouds in satellite imagery (e.g., see \cite{Welsch2004}, \cite{Welsch2009}, \cite{Liu2023}).

\cite{Welsch2004} provided a method called ILCT (Induction Local Correlation Tracking) for recovering photospheric velocities from magnetograms. The ILCT method is based on the induction equation, which relates the time derivative of the magnetic field to the curl of the electric field. By combining the induction equation with local correlation tracking (LCT), which is a method for estimating the velocity field of solar features on the solar surface, the ILCT method is able to obtain the velocity field from magnetograms. To use ILCT, the first step is to compute the curl of the electric field from the magnetogram data. This is done using the induction equation and assuming that the horizontal component of the electric field is small compared to the vertical component. The resulting curl of the electric field is then used as a proxy for the velocity field. Next, the ILCT method applies LCT to the curl of the electric field to obtain a smoother and more continuous velocity field. This involves dividing the curl of the electric field into smaller sub-images, on which LCT is applied separately. The resulting sub-velocity fields are then combined to obtain the full velocity field. The ILCT method has several advantages over other methods for estimating the velocity field from magnetograms. First, it is able to recover the full three-dimensional velocity field using only magnetogram data, which is important for understanding the dynamics of the solar corona. Second, it is able to obtain a smooth and continuous velocity field, which is important for accurately modeling the dynamics of the solar corona. Finally, it is computationally efficient, making it well-suited for analyzing large datasets. Overall, the ILCT method provides a powerful tool for studying the dynamics of the solar corona and for improving our understanding of the physical processes that govern the behavior of the Sun.

Another applicable technique is coherent structure tracking (CST), which assumes that the granules form a thermally coherent structure that naturally defines the underlying flow field. \cite{Roudier1999,Rieutord2007} used this method to determine the velocity profile of the flow. This algorithm involves five main steps as follows: image segmentation and granules detection, velocities measurement at the location of granule, velocity field reconstruction, computing field derivatives such as the z-component of the vorticity and divergence, noise estimation. The previous version of the algorithm was designed for a limited field of a few arcminutes, typically centered on the solar disk where solar rotation was removed through frame alignment. However, applying the CST to Solar Dynamic Observatory (SDO) / Helioseismic and Magnetic Imager (HMI) \cite{Schou2012} data requires a new granule time labeling method that accounts for the motion caused by solar rotation to avoid misidentification. In the previous version of the CST, time labeling was processed by tracking the barycenter of the granule. However, the presence of solar rotation in SDO/HMI data and the evolution of granules, such as the appearance of a new (small) granule in close proximity to an existing granule between two frames, can cause misidentification of barycenters. This results in poor temporal labeling and makes noise in the derived velocity maps. So, the new version of CST is improved by \cite{Roudier2012} to reconstruct the velocity field at scales larger than the sampling scale. Additionally, the updated version of CST provides the flexibility to select structures based on their characteristics, such as size, nature, and lifetime, thereby facilitating the study of their motion.

\begin{figure}
\centerline{\includegraphics[width=16cm,angle =90 ]{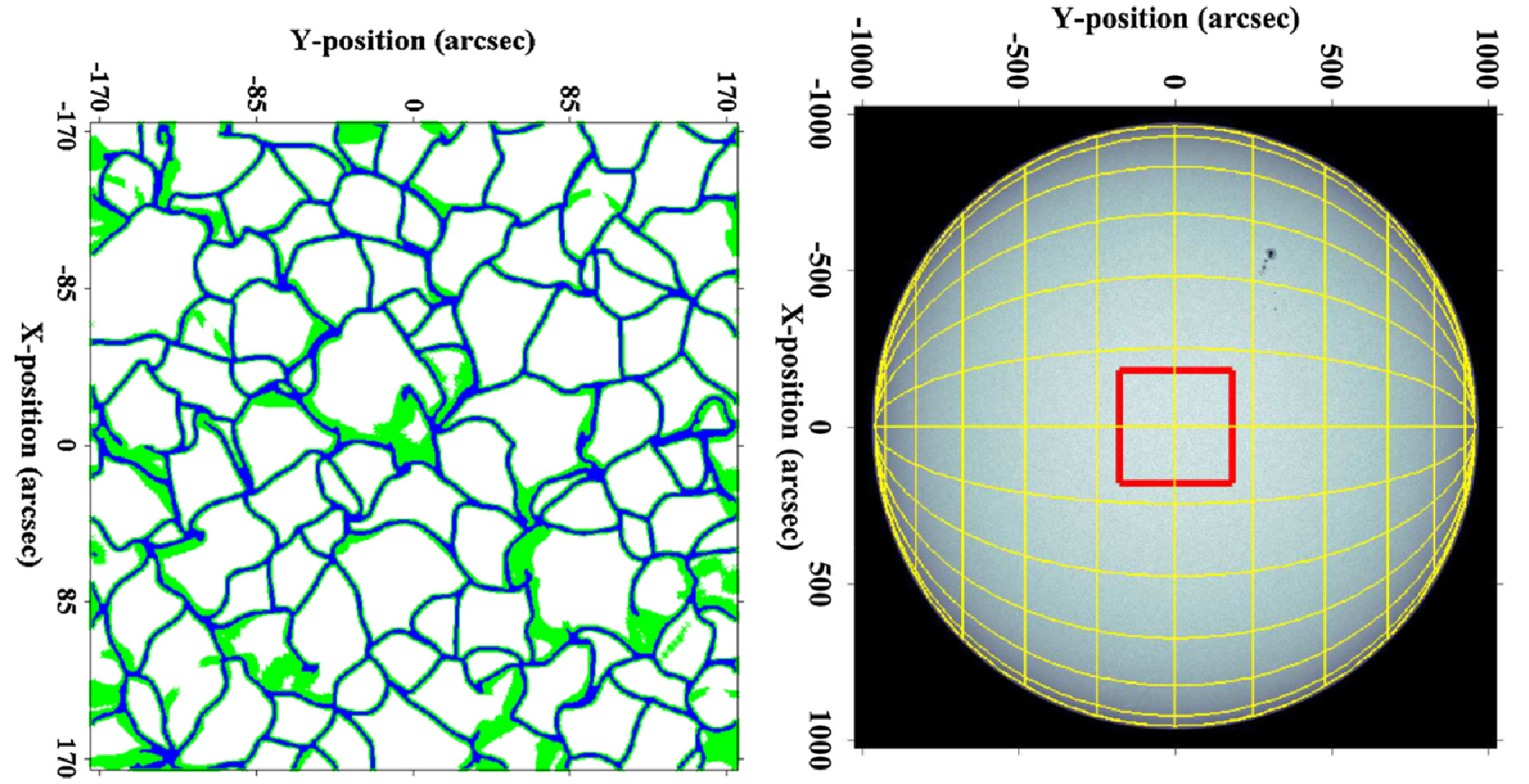}}
\caption[]{The full-disk image of the Sun captured on 30 December 2015 (00:00--00:30 UT) using the continuum SDO/HMI instrument, specifically at a wavelength of 6173 A, is displayed in the upper panel. Within this image, a red rectangular region is designated, encompassing an approximate area of 350$^{\prime\prime}$ $\times$ 350$^{\prime\prime}$. This region is centered around longitudes approximately $\pm$ 11 degrees around the central meridian, with latitude boundaries around $\pm$ 11 degrees around the solar equator. It's important to note that the utilization of equally-area projections such as Lambert and Postel projections has been omitted from this analysis. The lower panel showcases the outcome of the automated recognition technique, depicted in green. Overlaid on this panel, the blue lines denote the borders of individual cells, which have been derived through the application of morphological filters (further details are available in the accompanying text of the article \cite{Noori2019}.) Image reproduced with permission from \cite{Noori2019}, copyright by Elsevier.}
 \label{Majede-BT}
\end{figure}

\cite{Potts2004ball} proposed the ball-tracking (BT) method to track the flow of the solar photosphere using Solar and Heliospheric Observatory (SOHO) / Michelson Doppler Imaging (MDI) \cite{Scherrer1995} data. The BT used in the paper is based on the LCT method, which is a technique for estimating the velocity field of solar features on the solar surface. To use this technique, the first step is to apply LCT to each pair of magnetogram images to obtain a velocity field. The velocity field is then thresholded to identify regions of strong flow, which correspond to the boundaries of supergranules. Next, the BT method applies a morphological closing operation to the thresholded image to fill in gaps and smooth the boundary of the supergranules. The resulting image is then segmented using a watershed algorithm to identify individual supergranules. Finally, the BT method tracks the motion of each supergranule over time using a simple forward-backward tracking algorithm. This involves linking the boundaries of each supergranule in consecutive pairs of magnetogram images and calculating the displacement between the two boundaries. The displacement is then used to estimate the velocity of each supergranule. The BT method has several advantages over other methods for detecting and characterizing supergranules. First, it is able to track individual supergranules over time, which allows for a more detailed analysis of their properties. Second, it is able to accurately detect the boundaries of supergranules, even in the presence of noise and other artifacts. Finally, it is computationally efficient, making it well-suited for analyzing large datasets. It provides a powerful tool for studying the dynamics of supergranules and for improving our understanding of the physical processes that govern the behavior of the Sun. \cite{Attie2015MBT} developed BT for monitoring the evolution of individual magnetic features from magnetograms, which they call magnetic ball-tracking (MBT). This method enables the quantification of the flux of the tracked features and can track the footpoints of magnetic field lines inferred from magnetic field extrapolation. The algorithm is capable of detecting and quantifying both flux emergence and flux cancellation. A final output of BT method is presented in Figure \ref{Majede-BT}. The interested reader can refer to \cite{Attie2009}, \cite{Attie2015thesis}, \cite{Attie2016}, \cite{Yousefzadeh2016}, \cite{Attie2018}, and \cite{Noori2019} for more applications of BT.


The article \cite{DeForest2007} compares four different software codes for solar magnetic feature tracking: CURV, MCAT, SWAMIS, and YAFTA. The authors examine the differences between these codes in terms of their feature tracking behavior and parameterization, and make recommendations for best practices in future feature tracking work. The CURV code \cite{Strous1996} was the first code developed to study magnetic features in the SOHO/MDI quiet Sun data \cite{Hagenaar1999}, and is designed to identify and track individual magnetic features using a downhill search algorithm. The downhill search algorithm is based on the assumption that the magnetic features in a magnetogram are located on local minima in the magnetic field strength. The code tracks the features by moving downhill from one magnetogram to the next, following the features as they move and merge. The MCAT code \cite{Parnell2002} is designed to study the interaction between network flux elements, and uses a curvature-based algorithm to track magnetic features. The curvature algorithm is based on the assumption that magnetic features can be identified by the curvature of the magnetic field lines. The algorithm tracks features by identifying the regions of the magnetogram with the highest curvature, and following these regions over time. The SWAMIS code \cite{Lamb2003} is intended to drive semi-empirical MHD models of the quiet Sun, and uses a combination of the downhill and curvature tracking algorithms. The code identifies features using a downhill search, and then tracks them using the curvature algorithm. Finally, the YAFTA code \cite{DeForest2007} was developed to study active region dynamics, and uses a feature-based tracking algorithm that is based on the assumption that magnetic features can be identified by their shape and intensity. The algorithm tracks features by matching them to a library of templates that represent different types of magnetic features. In subroutines of the YAFTA, there are three main approaches used to identify structures of flux concentrations in solar magnetograms. The first algorithm is the clumping method, which considers pixels with fluxes above a threshold as a single feature \cite{Parnell2002}. The second algorithm is the downhill approach, which identifies one feature per local maximum region after thresholding \cite{Welsch2003}. This method can detect more patches than the clumping method. The third algorithm is the curvature method, which identifies the boundaries of features as a convex core around each local flux maximum \cite{Hagenaar1999}. However, the feature sizes obtained using the curvature method are smaller than with the other two approaches. Recent studies suggest that the downhill and clumping techniques produce more accurate segmentation results than the curvature method \cite{DeForest2007}.

The authors of the paper \cite{DeForest2007} compare the results of these approaches on a single set of data from SOHO/MDI, and identify the interplay between desired tracking behavior and parameterization tracking algorithms. YAFTA has certain limitations that can affect the accuracy of its results. One of these limitations is the use of a constant threshold value for segmenting and tracking magnetic features, which can cause some contiguous pixels to lose their connectivity conditions when their intensities fall below the threshold. This can result in the removal of these pixels from the segmented structures in consecutive frames, leading to errors in analysis due to the dynamic nature of the magnetic patches and the time lag between successive frames. Additionally, YAFTA has other potential shortcomings, such as the incorporation of previously-created variables with new ones after running the main program, which can lead to errors or program crashes if not renamed. Another issue is a bug in some architectures where Interactive Data Language (IDL) cannot concatenate newly-emerged magnetic structures with identical fields, as done in the CREATE-FEATURE subroutine. To address these bugs, \cite{Moradhaseli2021} modified the main program to associate the names of IDL structures with the numerators of each step.

\begin{figure}
\centerline{\includegraphics[width=15.2cm]{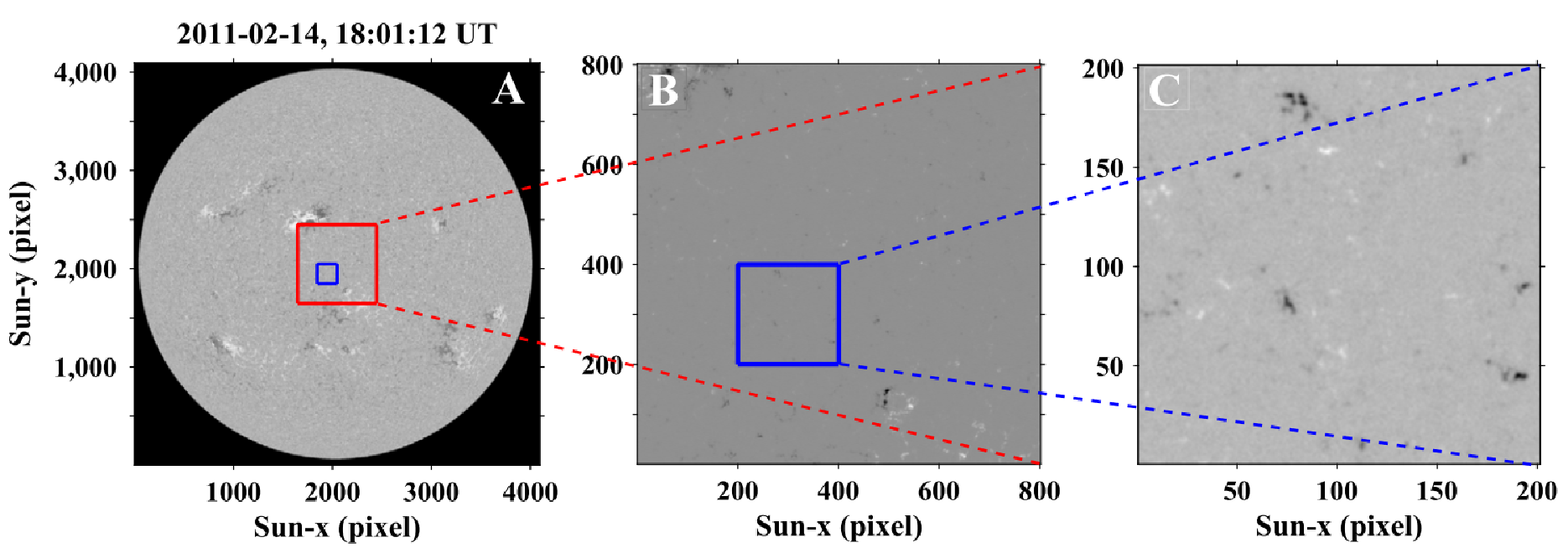}}
\caption[]{Here, there is a sample of SDO/HMI image with different crops employed to study the size distributions, flux distributions, and filling factors of positive and negative polarities in the QS based on YAFTA method. The Sun's full-disk magnetogram, as captured by SDO/HMI, was documented on 14 February 2011 at 18:01:12 UT (a). To extract the physical characteristics from both positive and negative elements, a cutout image from the solar equatorial region was selected (b). This cutout encompasses an area measuring 400$^{\prime\prime}$ by 400$^{\prime\prime}$ and was utilized for analysis throughout the year 2011, with a regular image acquisition rate of one image per day. For a specific investigation regarding the magnetic elements' physical attributes, especially their lifetimes, an image tile spanning 100$^{\prime\prime}$ by 100$^{\prime\prime}$ was cropped from all the acquired images (c). Related investigations focused on a three-day period, specifically from 14 to 16 February 2011, within the QS region. There was a time lag of 45 seconds between successive frames. More information can be find in \cite{Javaherian2017}. Image reproduced with permission from \cite{Javaherian2017}, copyright by Springer.}
 \label{HMI01}
\end{figure}

\begin{figure}
\centerline{\includegraphics[width=14cm]{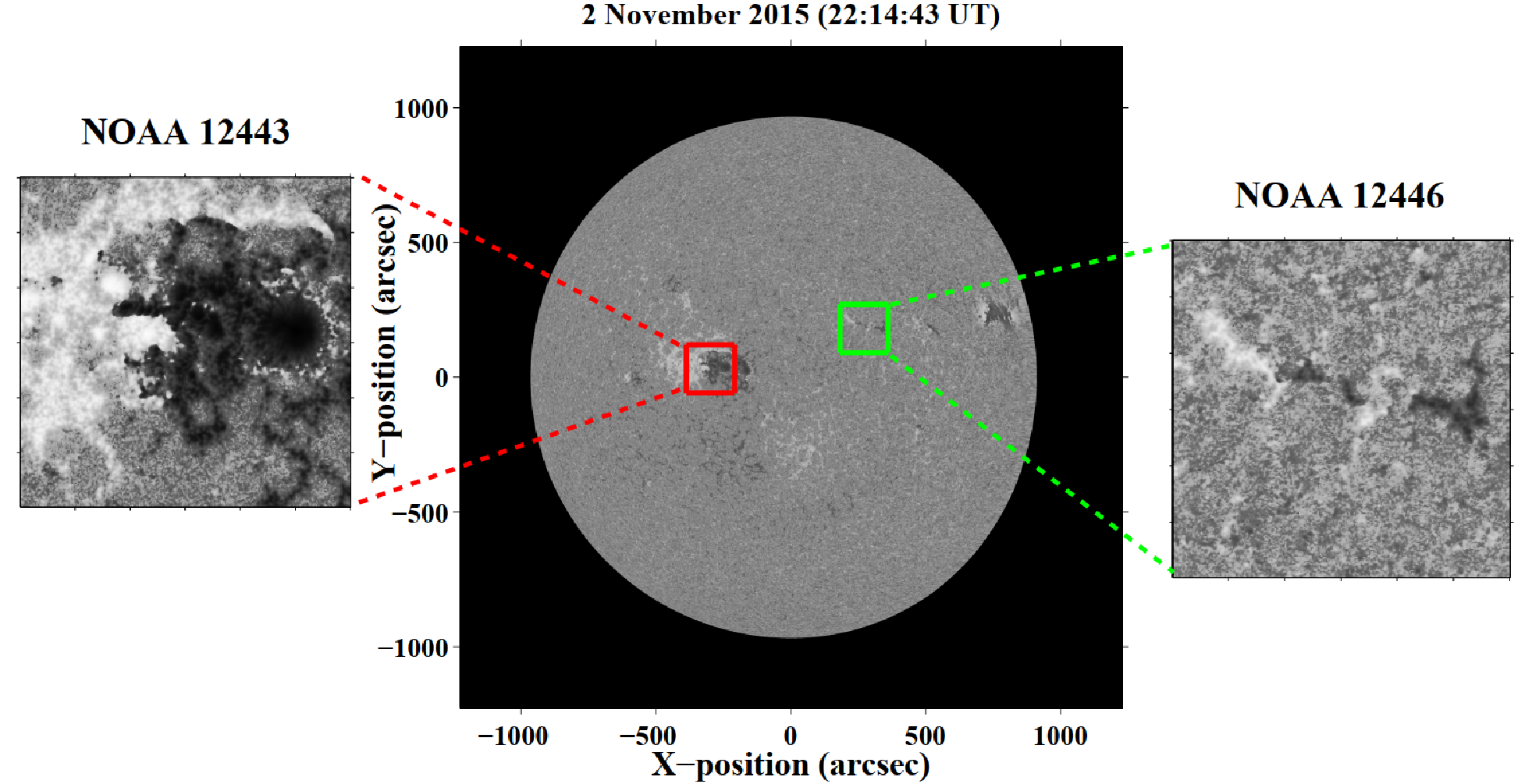}}
\caption[]{Here, there is another sample of SDO/HMI image with different crops employed to study the size distributions, flux distributions, and filling factors of positive and negative polarities inside both flaring and non-flaring ARs based on YAFTA method. An SDO/HMI magnetogram covering the entire solar disk was recorded on November 2, 2015. Delving into the details, the regions encompassing the ARs under scrutiny are outlined by contours with areas of 180$^{\prime\prime}$ by 180$^{\prime\prime}$. Among these contours, the one marked in red designates the flaring AR, which underwent analysis from November 3 to 5, 2015. In contrast, the green contour outlines the non-flaring AR, subject to examination between November 4 to 6, 2015. See more details in \cite{Moradhaseli2021}. Image reproduced with permission from \cite{Moradhaseli2021}, copyright by Acta Astronomica.}
 \label{HMI02}
\end{figure}

While \cite{Parnell2009} studied the flux distribution of magnetic elements, \cite{Javaherian2017} focused on other statistical parameters of patches, including size and lifetime frequencies. \cite{Javaherian2017} first used the YAFTA downhill algorithm to segment a large area of magnetograms from SDO/HMI data in 2011 (see Figure \ref{HMI01}, tiles A and B) and three days of quiet Sun data set (see Figure \ref{HMI01}, tiles B and C) including 5750 sequences. The algorithm was used for segmentation, but not tracking, and physical parameters such as size distribution, filling factors, and magnetic flux were extracted for both negative and positive polarities. In following, \cite{Moradhaseli2021} applied the YAFTA code to the constructed data cubes, which consisted of 5760 sequences of data spanning three days. To compare the statistical parameters of magnetic patches within flaring (Figure \ref{HMI02}, left panel) and non-flaring ARs (Figure \ref{HMI02}, right panel), two areas were selected from HMI magnetogram.

Various methods have been proposed for detecting sunspots using full-disk images. One common approach is the thresholding technique, used by several researchers like \cite{Preminger2001,Curto2008,Colak2008,Watson2009,ZhaoLin2016,Yang2018}. They applied this technique to data acquired from different observatories with various resolutions. \cite{Carvalho2020} examined results using thresholding and mathematical morphological operations on data received by the Coimbra Observatory. Other methods, such as Bayesian image-segmentation by \cite{Turmon2002,Turmon2010}, edge detection by \cite{Zharkov2005,Zharkov2006Zharkova}, and level-set image-segmentation by \cite{Goel2013}, were also used on MDI data. We attempted to explain the techniques of image processing used in these papers in details as follows:

\cite{Preminger2001} introduced a method based on contrast and contiguity in solar images. In order to successfully apply the described method, high-quality images are essential. The processing of CFDT1 images was carried out following the procedure outlined in Walton et al. (1998), with recent enhancements. Notably, faint vertical artifacts in CFDT1 images are now eliminated early in the processing, right after calibration and limb position fitting. This correction involves adjusting each column of the image to align its horizontal profile with its vertical profile's average. Additionally, CFDT1 images exhibit very low-level ghost images near the limb. To address this, a further correction is applied by multiplying each image with a correction factor computed from the median of about one year's worth of images at that specific wavelength. The images used for feature detection are contrast images obtained by dividing each image by an average limb darkening curve. The primary focus is on identifying long-lived magnetic features like faculae and sunspots, which influence solar irradiance. These features are expected to have a minimum size of a few pixels. The remainder of the solar disk is considered 'quiet Sun' with superimposed noise. Conventionally, the selection of features from a contrast image at a solar observatory involves a straightforward contrast criterion. This entails identifying pixels with contrasts that meet specific criteria for sunspots and faculae. However, this approach is not effective for detecting faint features as it can result in noise being mistaken for features. To address this, a novel technique has been developed. This technique identifies isolated features by scanning the disk and locating a pixel with a contrast exceeding a defined trigger criterion. The pixel is then used as a starting point, and neighboring pixels are examined to identify those forming a contiguous feature. Adjacent pixels meeting a less stringent contrast threshold criterion are considered part of the feature. This approach allows for more accurate detection of faint features by setting a low contrast trigger and requiring a minimum number of contiguous trigger pixels to validate the feature's presence. To ensure the effectiveness of this technique, probability calculations are employed. By considering the statistical properties of the quiet-Sun pixel distribution, suitable trigger and threshold contrast values are determined. Iteration between formal probability calculations and visual inspection is carried out to optimize the choices for trigger and threshold. Histogram analysis of the contrast image provides insights into quiet-Sun pixels. The distribution of quiet-Sun pixel contrasts is estimated by analyzing the negative contrast half of the histogram. For dark features, which are relatively scarce, a trigger contrast that yields a low false positive identification rate is selected. Conversely, for bright features, a slightly higher trigger contrast is chosen to accommodate the larger number of potential features while maintaining an acceptable false positive rate. The thresholds are set to values that ensure the selection of quiet-Sun pixels while minimizing false identifications.

The paper \cite{Turmon2002} explores image segmentation methods for extracting solar features from MDI data. The potential of the data is demonstrated through scatter plots generated from magnetograms and photograms taken 6 minutes apart. These plots allow distinguishing between sunspot umbra, penumbra, faculae, and quiet Sun. A focus is placed on these specific structures due to their direct identification in sample images. The procedure for analyzing such observations involves temporal and spatial adjustments of photograms, interpolation of magnetograms, and inference of labeling. The notation used involves feature vectors for pixels, which are part of images indexed by spatial coordinates. This article discusses the need for a uniform, automated technique with objective parameter determination. The process involves various steps, including temporal and flat-field corrections, interpolation, and labeling inference. The study emphasizes the application of these methods to capture solar structures in an objective and consistent manner.

The paper \cite{Zharkov2005} outlines an automated technique for detecting sunspots in SOHO/MDI white-light images. The technique employs edge detection and thresholding steps and is applied to a continuum image of the solar disk. The process involves aligning magnetogram data with the continuum image, identifying strong edges and high gradient regions using Gaussian smoothing and Sobel operator, iteratively thresholding and filtering candidate edges, detecting darker regions, and merging candidate maps. These steps produce a binary image with candidate regions containing sunspots. These regions are then refined using morphological operations and watershed functions. Sunspot detection within these regions involves further thresholding, considering quiet-Sun intensity and region statistics. The extracted parameters include umbral and penumbral characteristics, magnetic flux, and various observational data. The detected sunspots are stored in a catalog with search capabilities, facilitating future analysis and classification. The technique's potential for solar activity modeling and forecasting is also highlighted. The catalog, containing over 368,000 features from SOHO/MDI observations, is accessible online and through associated platforms.

In the context of image processing for solar images, \cite{Jones2007} discussed various methods for identifying and labeling specific features. These methods involve pre-processing the observations to eliminate artifacts and solar characteristics. One such method is the "one-trigger" (1T) approach, which applies simple thresholds to distinguish dark sunspots and bright faculae in different types of images. Another method, the "three-trigger" (3T) algorithm, utilizes more sophisticated contrast thresholds to identify larger feature areas with minimal confusion from instrumental noise. Additionally, a statistical method is employed, which avoids threshold usage and develops class-conditional models to produce probability images for sunspots, faculae, and QS. Another technique involves applying sequential thresholds to separate magnetized and unmagnetized regions and extract various features such as sunspots, ARs, and network structures. Thresholds are also applied to multidimensional histograms to isolate interesting subdomains, and factor analysis is used to establish spatial labeling based on the orthogonal combinations of observed subdomains. To compare the results obtained from different instruments, spatial registration techniques are employed, and the feature labelings are adjusted accordingly. Despite their differences, all these methods show comparable correlations with total solar irradiance and perform better for shorter periods.

\cite{Curto2008} presented a technique based on mathematical morphology to detect sunspots in full-disk solar images. The procedure for identifying and analyzing sunspots on the solar surface involves various steps aimed at image segmentation and characterization. The goal is to detect individual sunspots and group them into clusters based on their physical properties, contributing to determining the Wolf solar activity index. The entire process is facilitated by a graphic interface that allows operators to monitor and control the detection and grouping procedures. The results are published in solar bulletins and shared with the solar science community. The approach is executed as follows: $\bullet$ Image Segmentation and Detection of Individual Sunspots: The complex distribution of structures on the solar surface with varying intensity levels necessitates an image segmentation process. Three fundamental approaches for image segmentation are identified: boundary-based, region-based, and thresholding methods. In this context, the thresholding approach is chosen due to its simplicity, speed, and suitability for real-time processing. However, uneven brightness (limb darkening) and possible cloud shadows can complicate the histogram, making a global threshold insufficient. Therefore, pre-processing steps like image normalization and contrast enhancement are often applied. $\bullet$ Iterative Thresholding for Active Point Detection: The detection of individual sunspots involves finding active points, where the critical factor is identifying the appropriate threshold intensity level to distinguish active points from the background. An iterative thresholding method is employed. Initially, a low-intensity threshold is used to identify the darkest sunspot points. The threshold is then increased iteratively, counting the detected pixels at each step. The process continues until the detected pixel population stabilizes, indicating the background level has been reached. $\bullet$ Closing Transformation and Top Hat Operator: To identify smaller sunspots, a closing transformation is performed over the original image using a symmetric structuring element (SE). The SE size is iteratively increased to capture progressively larger sunspots. Subsequently, a Top Hat operator is applied to the closed image to detect points removed during the closing process. This step identifies fragments of sunspots and a sea of noisy background points. $\bullet$ Iterative Process for Sunspot Detection: The next phase involves iterative loops applied to the original image, one increasing the SE size and the other increasing the intensity level while controlling the growing population of sunspot pixels. This iterative approach captures various sizes of sunspots. $\bullet$ Region Growing for Sunspot Detection: An iterative process is used to assign each pixel belonging to the subset of true sunspot pixels to its corresponding sunspot. A region growing procedure groups pixels or subregions into larger regions based on neighborhood criteria. The process continues until all pixels have been assigned to sunspots. $\bullet$ Characterization of Sunspots: Characteristics of each sunspot, including position, mean intensity, and area, are computed and stored in an associated file. Both umbra and penumbra are considered a single unit in the sunspot, aligning with the former manual detection method. $\bullet$ Grouping of Sunspots: Sunspots are organized into groups that share physical properties. Each sunspot is assigned to a group based on criteria like proximity. This grouping process contributes to determining the Wolf solar activity index. $\bullet$ Verification and Quality Control: The results are crosschecked with synchronized solar magnetograms for verification. The neighborhood criterion is typically sufficient to classify sunspot groups. An operator supervising the process may redefine classifications in cases of very close group placements. $\bullet$ Completion of the Process and Reporting: At the conclusion of the process, a comprehensive list of individual sunspots and their respective groups is generated. This information is sent to a central solar data center and made available for publication in solar bulletins.

In study by \cite{Colak2008}, the process of sunspot detection and grouping involves several stages, including preprocessing, initial feature detection, and clustering. These steps are summarized as follows: $\bullet$ Preprocessing of MDI Images:
Preprocessing is divided into two stages: "Stage-1" processing for both intensitygram and magnetogram images involves solar disk detection, center/radius determination, solar coordinates calculation, and irrelevant information filtering. "Stage-2" processing, applicable only to magnetogram images, aligns images for correlation by resizing and rotation. $\bullet$ Initial Detection of Solar Features: Sunspots are detected using intensity filtering and region growing methods. The threshold value (Tf) is automatically calculated using mean and standard deviation. For active region detection, two threshold values are determined based on magnetic polarity. A region growing algorithm marks pixels within a window as active region candidates. $\bullet$ Deciding Active Regions and Grouping of Sunspots:
Active regions and sunspots are combined to cluster sunspots into groups. A process involving circular region marking, neural network application, and region coloration determines coupled polarity regions belonging to the same group.

In research provided by \cite{Watson2009}, image processing techniques are employed to automatically detect individual sunspots and establish new distributions of their appearances. This focus on individual spots, rather than groups recorded in existing catalogs, necessitates a simplified model of the Wilson effect. The sunspot identification algorithm leverages mathematical morphology tools, which involve erosion and dilation operations using a structuring element to analyze image features. The process starts by using SOHO/MDI continuum data, employing a morphological top-hat transform to enhance sunspot features. This involves eroding and dilating the image, effectively reducing sunspot size, then subtracting and thresholding to identify sunspot candidates. The algorithm is efficient and effective, capable of processing an image in seconds. It's tested against 'ground truth' images marked by human observers, demonstrating recovery of a significant portion of sunspot pixels while introducing a low rate of false positives. The transform is applied to a series of MDI white-light images over a period, and sunspot co-ordinates are determined and converted to heliographic co-ordinates. Comparing these with co-ordinates from subsequent images allows identification of consistent sunspots and tracking their longitudinal distribution over time. The results are compared with prior findings, and while some differences exist, statistical tests indicate consistency within a certain significance level.

The paper \cite{Goel2013} discusses a level-set method, specifically the Selective Binary and Gaussian Filtering Regularized Level Set (SBGFRLS) technique, for automated detection of sunspots from SOHO/MDI continuum images. This method uses statistical information both inside and outside the contours to create a region-based signed pressure force (SPF) function, which helps accurately locate sunspot boundaries even in weak or blurred edges. The SPF function guides the contour evolution to either shrink outside the object or expand within it. The level-set segmentation process involves initializing the Level Set Function (LSF), computing average intensities inside and outside the contour, evolving the LSF, regularizing it with a Gaussian function, and checking for convergence. The choice of the Gaussian filter's standard deviation ($\sigma$) and the evolution equation parameter ($\alpha$) significantly affects the results. This article also emphasizes an optimal LSF initialization strategy using a threshold intensity value. The technique was applied separately for umbra and penumbra detection, enhancing region extraction for sunspots. The method is illustrated with results from full-disk continuum images, demonstrating effective sunspot and pore detection.

The study \cite{ZhaoLin2016} introduces related tools for sunspot recognition. Mathematical morphology is employed for extracting image components based on shape and structure. Erosion and dilation operations, along with closing, opening, and Bot-hat transformation, are exploited. The Otsu algorithm \cite{Otsu1979,Sezgin2004}, used to find an adaptive threshold, is discussed next. Moving on to the recognition procedure, two steps are outlined. Firstly, the solar limb is extracted using morphological methods and Otsu algorithm. Secondly, sunspots are recognized within the solar limb using morphological Bot-hat operation and local threshold. Over-segmentation is reduced by imposing limits on sunspot properties. A procedure for recognizing sunspots in images with instrument noise is also presented.

The article \cite{Yang2018} employs SDO/HMI data to develop an automated sunspot detection method. SDO/HMI provides high-resolution solar white-light continuum images. Pre-processing involves exposure, dark current, flat field corrections, and cosmic-ray hit removal. About 13,800 images were selected for analysis. Limb-darkening is corrected by centering, creating an average radial profile, deriving a "Quiet Sun" background image, and obtaining a flat solar image. Sunspot detection employs normalized and smoothed images, followed by dual thresholds for umbra and penumbra segmentation. The process involves initializing parameters, evaluating fitness values, and performing genetic evolution with crossover and mutation. The final individual with the best fitness yields the optimal dual thresholds for distinguishing different solar regions.

The paper \cite{Carvalho2020} introduces an algorithm for automated sunspot detection using mathematical morphological transforms. The approach originated from attempts to describe geometric features in porous media and led to the development of mathematical morphology in image analysis. The technique involves comparing features to a structuring element, and it can be applied to both binary and grayscale images, offering versatility and quantitative measurements for complex shapes. The algorithm begins with pre-processing, involving operations like closing and opening to enhance the image. The process includes morphological reconstruction, adaptive threshold filtering, and addressing the circular nature of the solar disk. The algorithm then employs black top-hat transforms, adaptive thresholding, erosion, and reconstruction to identify sunspots and suppress noise. Umbra-penumbra segmentation is performed by analyzing histograms and applying thresholds. A comprehensive comparison with another method based on pixel intensity is also discussed. The morphological approach offers an automatic method for detecting sunspots and segmenting umbra-penumbra regions.

\section{Post-processing Procedures as Complementary Steps}

Post-processing procedures in solar image analysis play a crucial role in transforming the raw data and automated feature recognition results into meaningful and scientifically valuable insights. This stage involves converting numerical outputs, segmented images, datacubes, and movies into formats that are suitable for communication, further analysis, and theoretical modeling. This explanation will delve into the various post-processing tasks outlined in the provided text and provide extended insights, along with relevant references.

Visualization of Solar Imagery: Visual representation of solar data has evolved from simple drawings to complex multi-dimensional displays. As solar observations have become more sophisticated, the need for advanced visualization techniques has increased. Software tools have emerged to meet these needs, offering capabilities such as overlaying images, creating composite images, generating movies, and constructing 3D reconstructions. Examples of such software include SolarSoft (SSW) \cite{Freeland1998}, SolarMonitor, SolarWeather Browser (SWB) \cite{LeGall2008}, and FESTIVAL \cite{Nicula2008}. Each software package has specific advantages and specialties, contributing to the overall visualization landscape.

Cataloguing of Solar Features: The vast amount of solar data necessitates efficient cataloging and metadata organization. Feature and event catalogs facilitate data analysis by enabling searches based on well-defined parameters. Various virtual solar observatories, such as the Virtual Solar Observatory (VSO) \cite{Hill2004}, SDO Heliophysics Event Knowledgebase (HEK) \cite{Hurlburt2010,Somani2010}, and European Grid of Solar Observatory (EGSO) \cite{Bentley2002a,Bentley2002b}, provide tools to interconnect databases and enable user-friendly queries.

Statistics of Solar Events: Statistical analysis of solar events, such as sunspots \cite{Walton2003,Imada2020}, photospheric granules \cite{Javaherian2014} and BPs \cite{Feng2013}, coronal BPs \cite{Alipour2015}, CHs \cite{Tajik2023}, ARs \cite{Arish2016}, solar flares, microflares, and nanoflares (see \cite{Hudson1991,Yashiro2006,Tajfirouze2012} and references  therein) is crucial for understanding their occurrence and behavior. Image processing techniques play a pivotal role in ensuring unbiased and accurate event segregation and measurement. Various statistical studies rely on automated algorithms to detect events across different scales, eliminating human subjectivity and bias. For instance, the power-law distribution of flare energies and the correlation between flare and CME energies require accurate and objective event detection.

Theoretical Modeling: Theoretical modeling of solar physical processes often relies on automated feature detection, as these models require precise measurements of features in both space and time. Automated detection of features, such as EUV loops, is essential for 3D reconstruction modeling \cite{Aschwanden2005}, magnetic field extrapolation \cite{Aschwanden2008SoPh}, stereoscopy-based tomography \cite{Barbey2008}, and other advanced techniques (e.g., \cite{Koning2009}). Automated pattern recognition ensures high precision and objectivity, which are critical for accurate modeling.

Prediction and Forecasting: Real-time forecasting in solar physics heavily relies on automated pattern recognition due to the need for rapid analysis and decision-making. Automated detection of features, combined with probabilistic prediction algorithms and machine learning, forms the basis of automated forecasting systems. These systems are employed for predicting various solar events, such as flares, CMEs, and filament eruptions (e.g., see \cite{Gallagher2002,Qahwaji2007,Olmedo2008,Alipour2019} and references therein).

Confirmation or Rejection of Previous Results: In the scientific process, the confirmation or rejection of previous results or theories often hinges on the acquisition of new data and evidence. This process is fundamental to the advancement of scientific understanding and knowledge. Here's how new results obtained from data can lead to the confirmation or rejection of previous results or theories in science. For this purpose, we itemized the scientific process as follows: $\bullet$ Hypotheses and Theories: Scientific research begins with hypotheses or theories that attempt to explain natural phenomena. These hypotheses are based on existing knowledge and observations. $\bullet$ Data Collection and Experimentation: To test these hypotheses, scientists design experiments, observations, or gather data through various methods. These methods are chosen to be as unbiased and controlled as possible to ensure the accuracy and reliability of the data collected. $\bullet$ Comparison with Existing Knowledge: The newly acquired data is then compared to existing knowledge, theories, and predictions. If the data aligns well with the predictions made by the existing theory, it provides confirmation of those theories. This doesn't necessarily prove the theory true, but it does provide stronger support for its validity. $\bullet$ Consistency and Reproducibility: Confirmation of previous results occurs when other scientists can reproduce the same results independently using the same methods. This adds to the reliability of the findings. $\bullet$ Deviation and Inconsistencies: Sometimes, new data might show inconsistencies with existing theories or predictions. These inconsistencies can lead to further investigation to understand why the observed results deviate from expectations. $\bullet$ Re-evaluation and Modification: When new data contradicts existing theories, scientists might re-evaluate the theories. They might modify the theories to accommodate the new data or revise them entirely. This process of revising theories in response to new evidence is a cornerstone of scientific progress. $\bullet$ Rejection of Hypotheses or Theories: If new data consistently and robustly contradicts the predictions of a theory, it might lead to the rejection of that theory. However, this rejection is not taken lightly; it requires a substantial body of evidence that consistently undermines the theory. $\bullet$ Iterative Process: Science is an iterative process. New data can lead to adjustments in theories, which in turn generate new hypotheses to test. This cycle of proposing, testing, and refining theories drives the advancement of knowledge. So, the acquisition of new data is a crucial element in the confirmation or rejection of previous results or theories in science and solar physics. Data that aligns with existing theories provides confirmation, while data that deviates can lead to modifications or rejections of theories. This constant interplay between theory and data drives scientific progress by refining our understanding of the natural world. In recent works of solar physics, the graph-based approaches in a framework of complex system were introduced to confirm the previous results obtained by the other methods. As an example, \cite{Lotfi2020} exploited the unsupervised method in a framework of complex network to provide a nw approach for diagnosis of flares in ultraviolet emission band (1600 \AA). The obtained results had consistency with events detected by Geostationary Operational Environmental Satellite (GOES).


Comparison with Simulation: In solar physics, two common types of simulations used to study the behavior of the Sun are $N$-body simulations and grid-based simulations. Here are the key differences between these two approaches: $\bullet$ Particle Representation: $N$-body simulations model individual particles, such as electrons, protons, or dust grains, as discrete entities that interact with each other through gravitational or electromagnetic forces. In contrast, grid-based simulations divide the simulated region into a grid or mesh, where physical properties like density, temperature, and velocity are assigned to each grid cell. $\bullet$ Spatial Resolution: $N$-body simulations typically have higher spatial resolution compared to grid-based simulations. Since $N$-body simulations directly model individual particles, they can capture fine-scale details and interactions. On the other hand, grid-based simulations sacrifice some spatial resolution due to the discretization of the simulated region into grid cells. $\bullet$ Complexity and Flexibility: Grid-based simulations are more versatile and can handle a wider range of physical processes. They can incorporate complex physics, such as radiative transfer, magnetohydrodynamics, and energy transport mechanisms. N-body simulations, while simpler in terms of their particle interactions, may not be able to capture these additional physical processes. $\bullet$ Computational Efficiency: $N$-body simulations can be computationally expensive, especially when simulating a large number of particles or when including complex interactions. On the other hand, grid-based simulations are generally more computationally efficient since they operate on a fixed grid structure, allowing for faster calculations and easier parallelization. $\bullet$ Applicability: $N$-body simulations are commonly used in scenarios where individual particle interactions are crucial, such as modeling the dynamics of charged particles in the solar wind or the motion of dust grains in the Sun's environment. Grid-based simulations, on the other hand, are well-suited for studying large-scale phenomena like the behavior of the Sun's magnetic field, convective motion, or the propagation of waves through the solar atmosphere.

In fact, $N$-body simulations focus on modeling individual particle interactions and are suitable for scenarios where detailed particle-level dynamics are important (e.g., see \cite{Rasio2000,Aarseth2003,Springel2005,Yousefzadeh2021}. Grid-based simulations, on the other hand, provide a more versatile and computationally efficient approach for studying large-scale phenomena and incorporating complex physical processes in solar physics (e.g., see \cite{Aschwanden2004,Gudiksen2005,Vogler2007,Javaherian2016}. Both simulation methods have their strengths and limitations, and the choice between them depends on the specific research question and the level of detail required.

There are so many comparisons between different types of simulations used in solar physics and data. As a note, we list some of the few examples of simulations compared with results extracted from data: $\bullet$ MHD Simulations: MHD simulations are widely used in solar physics to study the behavior of the Sun's plasma, which is a combination of charged particles and magnetic fields. These simulations help model phenomena like solar flares, coronal mass ejections (CMEs), and the generation and evolution of the Sun's magnetic field. MHD simulations allow scientists to investigate the complex interplay between plasma dynamics and magnetic fields in the Sun's atmosphere. $\bullet$ Radiative Transfer Simulations: They are used to understand the interaction of radiation with the Sun's atmosphere. These simulations help model how energy is transported through the layers of the Sun, including the photosphere, chromosphere, and corona. By simulating the absorption, emission, and scattering of radiation, scientists can study the temperature, density, and composition of different solar atmospheric layers. $\bullet$ Solar Interior Simulations: Simulations of the solar interior aim to model the physical processes occurring within the Sun's core and radiative and convective zones. These simulations help understand how energy is generated and transported within the Sun, as well as the dynamics of convection and the formation of sunspots. By simulating the interior of the Sun, scientists can gain insights into its structure, evolution, and the mechanisms driving solar activity. $\bullet$ Solar Dynamo Simulations: Dynamo simulations focus on modeling the generation and evolution of the Sun's magnetic field. These simulations help understand the processes that give rise to the Sun's magnetic activity, such as the formation of sunspots, solar cycles, and the reversal of the Sun's magnetic poles. By simulating the dynamo processes, scientists can investigate the mechanisms responsible for the Sun's magnetic behavior. $\bullet$ Solar Wind Simulations: Solar wind simulations aim to model the flow of charged particles from the Sun into space. These simulations help study the properties and dynamics of the solar wind, including its speed, density, and magnetic field structure. By simulating the solar wind, scientists can gain insights into its impact on space weather and its interaction with planetary magnetospheres. These are just a few examples of the different types of simulations used in solar physics.

Post-processing procedures in solar image analysis encompass a wide range of tasks that transform raw data and automated feature recognition results into valuable scientific insights. These tasks include visualization, cataloging, statistical analysis, theoretical modeling, prediction, confirmation of previous results, and comparison with simulations. Image processing methods plays a critical role in ensuring objectivity and accuracy throughout these post-processing stages, ultimately contributing to a more comprehensive understanding of solar phenomena and their impacts on space weather and astrophysical processes (e.g., see \cite{Aschwanden2010a,Aschwanden2010b} and references therein).

\section{Conclusion}

The rapid growth of data, especially in recent years, has led to a greater need for processing data in all areas of science. Owing to the continuous recording of data in the field of astronomy, and subsequently in its sub-branches such as solar physics, data processing assumes a crucial role in comprehending interconnected concepts. This is largely due to the combination of space missions and telescopes that observe the Sun, which provide high-quality images and data. These efforts have improved our understanding of the Sun's behavior and its effects on Earth. This paper serves as a source of knowledge about solar photospheric observations, image processing, and the journey from raw data to meaningful results. It shows how science and tools work together to uncover the Sun's mysteries and its relationship with Earth.

This paper has highlighted various ways to analyze images of the Sun's surface. We started by explaining what the Sun's surface is like, which helps set the stage for the rest of the paper. We then looked into preparing the images before analysis, an important step. Additionally, we reviewed studies that used image processing techniques to study the Sun, both in the past and with new methods. This shows how technology is shaping future research. Lastly, we explored how to enhance results from raw data through post-processing. These steps, though less obvious, are vital for turning data into useful insights. By explaining these processes, this paper aims to be a comprehensive guide for researchers, helping them make sense of raw data and draw meaningful conclusions.

\section*{Acknowledgment}
The German contribution to Sunrise is funded by the Bundesministerium $\rm f\ddot{u}r$ Wirtschaft und Technologie through the Deutsches Zentrum $\rm f\ddot{u}r$ Luft- und Raumfahrt e.V. (DLR), Grant No. 50 OU 0401, and by the Innovationsfonds of the President of the Max Planck Society (MPG). The Spanish contribution has been funded by the Spanish MICINN under projects ESP2006-13030-C06 and AYA2009-14105-C06 (including European FEDER funds). HAO/NCAR is sponsored by the National Science Foundation, and the HAO Contribution to Sunrise was partly funded through NASA grant number NNX08AH38G.

The authors thank the Editorial Board of Iranian Journal of Astronomy and Astrophysics for giving us the opportunity of reviewing "Image Processing Methods in Solar Photospheric Data Analyzes" on their Special Issue about the Sun.

\end{document}